\documentclass[showpacs,preprintnumbers]{revtex4}
\usepackage{amssymb}
\usepackage{amsmath}
\usepackage{graphicx}
\usepackage{dcolumn}
\usepackage{bm}
\usepackage{subfigure}
\usepackage[colorlinks,citecolor=blue, linkcolor=blue,hyperindex,dvipdfm]{hyperref}

\begin{document}

\title{Entropic force between two horizons of dilaton black holes with a power-Maxwell field}

\author{Hui-Hua Zhao$^{1}$, Li-Chun Zhang$^{1}$\footnote{e-mail:zhlc2969@163.com(L.-C. Zhang), corresponding author}, Ying Gao$^{2}$, Fang Liu$^{1}$}

\affiliation{$^{1}$ Institute of Theoretical Physics, Shanxi Datong University, Datong, 037009, China
\\ $^{2}$ School of Mathematics and Statistics, Shanxi Datong University, Datong, 037009, China}

\begin{abstract}
In this paper, we consider $(n+1)$-dimensional topological dilaton de Sitter black holes with power-Maxwell field as thermodynamic systems. The thermodynamic quantities corresponding to the black hole horizon and the cosmological horizon respectively are interrelated. So the total entropy of the space-time should be the sum of the entropies of the black hole horizon and the cosmological horizon plus a corrected term which is produced by the association of the two horizons. We analyze the entropic force produced by the corrected term at given temperatures, which is affected by parameters and dimensions of the space-time. It is shown that the change of entropic force with the position ratio of two horizons in some region is similar to that of Lennard-Jones force with the position of particles. If the effect of entropic force is similar to that of Lennard-Jones force, and other forces are absent, the motion of the cosmological horizon relative to the black hole horizon would have an oscillating process. The entropic force between the two horizons is probably one of the participants to drive the evolution of universe.

\textbf{Keywords}: entropy, entropic force, dilaton dS space-time
\end{abstract}

\pacs{04.70.-s, 05.70.Ce}

\maketitle

\section{Introduction}
In the early cosmic inflation, our universe is a quasi-asymptotic de Sitter (dS) space-time, where the introduced cosmological constant term can be seen as the vacuum energy. If the cosmological constant corresponds to dark energy, our universe will evolve into a new de Sitter phase. In order to construct the whole evolutionary history for our universe and find out the reason of the accelerated expansion as well, the classical, quantum and thermodynamic properties of dS space-time should be studied. In addition, the success of the correspondence between Anti-de Sitter space and conformal field theory (AdS/CFT) has prompted people to find a similar dual relationship for de Sitter space-time.

In recent years, the thermodynamic properties and possible phase transition of dS black holes have been studied extensively  \cite{Saoussen19,Fil19,Brian13a,Hendi17a,Fil18,Pappas16,Kanti17,Sekiwa06,Kubizanak16,McInerney16,Azreg-Ainou15b,ZhangLC16a,Dinsmore19,Guo15,Zhao14b,Urano09,Sourav13,Cai02a,Cai02b,Cai02c,Arindam19,ZhangLC16b}. For a dS black hole there are a black hole horizon and a cosmological horizon, and, in general, the radiation temperatures of the two horizons are different. Considering the two horizons as two thermodynamic systems they satisfy the first law of black hole thermodynamics respectively, but their thermodynamic quantities are interrelated because of their common quantities, mass, electric charge, and cosmological constant. In most of the previous works on thermodynamic properties of dS black holes, the whole entropy of dS black hole is seen as the sum of the entropies of black hole horizon and cosmological horizon. Considering the correlation of the two horizons, a corrected term for the whole entropy is required, which is derived and analyzed in this work.

In 2011, Verlinde \cite{Verlinde11} thought of linking gravity to an entropic force. Gravity emerges as a consequence of information regarding the positions of material bodies, combining a thermal gravitation treatment to 't Hooft's holographic principle. The ensuing conjecture was later proved \cite{Plastino18,Zamora18,Panos19,Calderon19,Plastino19a,Plastino19b,Plastino19c,Plastino19d} in a classical scenario.  Accordingly, gravitation ought to be viewed as an emergent phenomenon. Such exciting Verlinde's idea received a lot of attention\cite{Nobuyoshi19,Nobuyoshi17,Nobuyoshi16,Cai11a,Cai10a,Treumann19}. So the entropic force is an important force in the universe and it is probably one of the participants to drive the cosmic accelerated expansion.

As an explanation for the cosmic accelerated expansion, the early theory of dark energy has been proposed by Riess \cite{Rezaei19,Riess11,Riess19,Riess98}. In this theory, the cosmic accelerated expansion is caused by an exotic component called dark energy, which accounts for about $73{\rm{\% }}$ of the universe's capacity according to astronomical observations. Astronomers assume that dark energy exists in the first second after the Big Bang. the Big Bang pushes all matter to the whole space, and then the initial expansion begins. Shortly after the Big Bang, dark energy bumps several times, which causes the present cosmic accelerated expansion. Some models for dark energy evolution have been proposed. For example, if the equation of state of dark energy is $P = \omega \rho $, where $\omega $ is the parameter of the equation of state, the evolution of dark energy is exponential with the power $3(1 + \omega )$ when $\omega  >  - 1$ \cite{Peebles88,Peebles03,Chevalier01,Toledo19,Komatsu19}. The effect of dark energy is to make the cosmic accelerated expansion\cite{Nobuyoshi19}, which means it has some kind of repulsion. But what is the nature of dark energy? How does it evolve with the cosmic expansion? So far people don't really know.

In addition, the possible reason for the cosmic accelerated expansion is dark matter, which interacts more strongly with normal matter or radiation  than previously assumed. The existence of dark matter in the universe has been a common sense of modern cosmology. Dark matter accounts for about $23\%$ in the total cosmic components. Dark matter does not participate in electromagnetic interaction, nor interact with photons. The latest research shows \cite{Tenkanen19a,Tenkanen19b,Tenkanen19c} that dark matter appeared earlier than normal matter during the expansion of the universe, although normal matter is produced during the Big Bang. A kind of non-spin scalar particle is produced during the rapid cosmic expansion. Up to now, only one type of scalar particle has been found, which is the famous Higgs boson. According to Tenkanen, it can be the candidate of dark matter. But how do the dark energy, the dark matter and the total energy of the universe evolve in the expansion universe? The real reason for the cosmic accelerated expansion are still not clear.

The fact that the Universe expands with acceleration along the scheme of standard Friedmann model\cite{Riess98} created much more interest in the alternative theories of gravity in recent years, one of which is dilaton gravity. Dilaton gravity can be thought as the low energy limit of string theory, and one recovers Einstein gravity along with a scalar dilaton field, which is nonminimally coupled to the gravity and other fields such as gauge fields. We are interested in studying the properties of the dilaton black holes when the gauge field is in the form of the power-Maxwell field\cite{Dayyani17}. Being different from the linear electromagnetic field, nonlinear electrodynamics was introduced to remove the central singularity of the point-like charges and obtained finite energy solutions for particles by extending Maxwell's theory. In cosmology, one can call upon nonlinear electrodynamics to explain the inflationary epoch and the late-time accelerated expansion of the universe\cite{Garcia-Salcedo2000,Novello2004}. A variety of nonlinear electrodynamics models\cite{Soleng1995,Hassaine2007,Hassaine2008,Hendi2012,Hendi2013,Kruglov2015} have been proposed and studied extensively, in which the power-Maxwell field\cite{Hassaine2007} is conformally invariant in $(n+1)$-dimensional space-time for $p=(n+1)/4$, where $p$ is the power parameter of the Power-Maxwell Lagrangian, while the Maxwell Lagrangian is only conformally invariant in four dimensions.

The effect of the dilaton field \cite{Dehghani14,Zhao13,Zhao14b}and the power-law Maxwell field\cite{Zangeneh15a,Hendi15,Zangeneh15b,Hendi13} on thermodynamics of anti-de Sitter (AdS) black holes has been studied in extended phase space. However, as far as we know, the discussion for the effective thermodynamic quantities with effects of the power-law Maxwell field and the dilaton field has not been done in de Sitter space-time. In this paper, we study entropy and entropic force of dilaton black holes coupled to nonlinear power-Maxwell field in de Sitter space-time, and investigate the effects of exponent $p$, the dilaton coupling constant $\alpha$ and the space-time dimension $n$ on the entropy and the entropic force of the black holes in dS space-time and to explore their connection to the expansion of the universe.

This paper is organized as follows. The solutions of charged dilaton black hole with power-Maxwell field in dS space-time are introduced in section 2. The thermodynamic quantities respectively corresponding to the black hole event horizon and the cosmological envent horizon in the dS space-time are given in section 3. Considering the correlation of two horizons the effective thermodynamic quantities and the modified entropy are derived in section 4. In section 5, according to the relationship between entropy and entropic force, the entropic force between the two horizons is obtained and analyzed. Conclusions and discussions are given in the last section. The units ${G_{n + 1}} = \hbar  = {k_B} = c = 1$ will be used throughout this work.

\section{Black hole solution of Einstein power-Maxwell-dilaton fields in dS Space-time}

In this section, we introduce the action of Einstein power-Maxwell-dilaton (EPMD) gravity, the solutions of the EPMD field equations, and the mass, the electric charge, and the cosmological constant of the EPMD black holes in dS space-time.

The action of $(n+1)$-dimensional $(n\ge3)$ EPMD gravity can be written as \cite{Dayyani17,Dayyani18,Dehghani16,Dehghani14,Mo16}

\begin{equation}  \label{2.1}
I =  - \frac{1}{{16\pi }}\int {{d^{n + 1}}} x\sqrt { - g} \left[ {R - \frac{4}{{n - 1}}{{(\nabla \Phi )}^2} - V(\Phi ) + {{(-{e^{ - 4\alpha \Phi /(n - 1)}}{F_{\mu \nu }}{F^{\mu \nu }})}^p}} \right],
\end{equation}
which can yield the follow field equations by taking the action as varying with respect to the gravitational field $g_{\mu,\nu}$, dilaton field $\phi$ and the gauge field $F_{\mu,\nu}$,
\begin{equation}  \label{2.2}
{R_{\mu \nu }} = \left[ {\frac{1}{{n - 1}}V(\Phi ) + \frac{{2p - 1}}{{n - 1}}{{\left( { - F{e^{ - 4\alpha \Phi /(n - 1)}}} \right)}^p}} \right]{g_{\mu \nu }} + \frac{4}{{n - 1}}{\partial _\mu }\Phi {\partial _\nu }\Phi  + 2p{e^{ - 4\alpha p\Phi /(n - 1)}}{( - F)^{p - 1}}{F_{\mu \lambda }}{F_\nu }^\lambda ,
\end{equation}

\begin{equation}  \label{2.3}
\nabla ^{2}\Phi  - \frac{{n - 1}}{8}\frac{{\partial V}}{{\partial \Phi }} - \frac{{\alpha p}}{2}{e^{ - 4\alpha \Phi /(n - 1)}}{( - F)^p} = 0,
{\partial _\nu }\left[\sqrt{-g} {{e^{ - 4\alpha \Phi /(n - 1)}}{{( - F)}^{p - 1}}{F^{\mu \nu }}} \right] = 0,
\end{equation}
where $R$ is the Ricci scalar, $V(\Phi )$ is a potential for  $\Phi $, $p$ and $\alpha $ are two  constants determining the nonlinearity of the electromagnetic field and the strength of coupling of the scalar and electromagnetic field, respectively.  ${F_{\mu \nu }} = {\partial _\mu }{A_\nu } - {\partial _\nu }{A_\mu }$ is the electromagnetic field tensor and ${A_\mu }$ is the electromagnetic potential.

The topological black hole solutions take the form \cite{Dayyani17,Dayyani18,Dehghani16,Dehghani14,Mo16}
\begin{equation}  \label{2.4}
d{s^2} =  - f(r)d{t^2} + \frac{{d{r^2}}}{{f(r)}} + {r^2}{R^2}(r)d\Omega _{n - 1}^2,
\end{equation}
where
 \begin{equation}  \label{2.5}
  f(r) =  - A{r^{2\gamma }} - \frac{m}{{{r^{(n - 1)(1 - \gamma ) - 1}}}} + {q^{2p}}B{r^{ - {\textstyle{{2[(n - 3)p + 1] - 2p(n - 2)\gamma } \over {2p - 1}}}}} + C\Lambda {r^{2(1 - \gamma )}},
  \end{equation}
and $A = \frac{{k(n - 2){{({\alpha ^2} + 1)}^2}{b^{ - 2\gamma }}}}{{({\alpha ^2} - 1)({\alpha ^2} + n - 2)}}$,\quad $B = \frac{{{2^p}p{{({\alpha ^2} + 1)}^2}{{(2p - 1)}^2}{b^{ - {\textstyle{{2(n - 2)p\gamma } \over {(2p - 1)}}}}}}}{{\Pi ({\alpha ^2} + n - 2p)}}$,\quad$C = \frac{{2{{({\alpha ^2} + 1)}^2}{b^{2\gamma }}}}{{(n - 1)({\alpha ^2} - n)}}$,
in which $b$ is an arbitrary nonzero positive constant, $\gamma  = {\alpha ^2}/({\alpha ^2} + 1)$ , $\Pi  = {\alpha ^2} + (n - 1 - {\alpha ^2})p$ .

Note that $\Lambda $  remains as a free parameter and ${\Lambda}>0$ in dS space-time, and it plays the role of the cosmological constant

 \begin{equation}  \label{2.7}
 \Lambda  = \frac{{(n - {\alpha ^2})(n - 1)}}{{2{l^2}}},
 \end{equation}
where $l$ denotes the ADS length scale. In the Eq.(\ref{2.5}), $m$ appears as an integration constant and is related to the ADM (Arnowitt-Deser-Misnsr) mass of the black hole. According to the definition of mass due to Abbott and Deser \cite{Abbott82,Olea05}, the mass of the solution Eq.(\ref{2.5}) is \cite{Sheykhi07}

 \begin{equation}  \label{2.8}
 M = \frac{{{b^{(n - 1)\gamma }}(n - 1)}}{{16\pi ({\alpha ^2} + 1)}}m,
 \end{equation}

The electric charge $Q$ and potential $U$ are expressed as\cite{Dayyani17}
 \begin{equation}  \label{2.9}
 Q = \frac{{{2^{p - 1}}{q^{2p - 1}}}}{{4\pi }},  U = \frac{{(n - 1){p^2}q{b^{{\textstyle{{(2p - n + 1)\gamma } \over {2p - 1}}}}}}}{{\Pi \Upsilon r^\Upsilon }},  \Upsilon  = \frac{{n - 2p + {\alpha ^2}}}{{(2p - 1)(1 + {\alpha ^2})}}.
 \end{equation}

The fact that the electric potential $U$ should have a finite value at infinity and the term including $m$ in the solution $f(r)$ in
spacial infinity should vanish lead to the restrictions on $p$ and $\alpha$\cite{Zangeneh15a}.
 \begin{equation}  \label{res1}
 \frac{1}{2}<p<\frac{n+\alpha ^{2}}{2}.
 \end{equation}
 \begin{equation}  \label{res2}
 \alpha ^{2}<n-2.
 \end{equation}

\section{Thermodynamic quantities of the two horizons of the EPMD black holes in dS Space-time}

The dS black holes have two horizons, which are BEH and CEH. The BEH locates at $r = {r_ + }$ and the CEH locates at $r = {r_c}$. The positions of them can be determined by $f({r_ + }) = 0$ and $f({r_c}) = 0$ respectively. Thermodynamic quantities on BEH and CEH satisfy the first law of thermodynamics respectively \cite{Dinsmore19,Brian13a,Sekiwa06}. In this section, we introduce the thermodynamic quantities corresponding to the BEH and CEH respectively. Replace the $r$ in the Eq.(\ref{2.9}) with $r_+$ or $r_c$, one can get the electric potentials of BEH or CEH.

The surface gravities of BEH and CEH are respectively given by
\begin{equation}  \label{2.10}
{\kappa _ + } = \frac{1}{2}{\left. {\frac{{df(r)}}{{dr}}} \right|_{r = {r_ + }}} = \frac{{(1 + {\alpha ^2})}}{2}\left( {\frac{{k(n - 2){b^{ - 2\gamma }}}}{{(1 - {\alpha ^2})}}r_ + ^{2\gamma  - 1} - \frac{{2\Lambda {b^{2\gamma }}}}{{n - 1}}r_ + ^{1 - 2\gamma } - \frac{{{2^p}p(2p - 1){b^{ - {\textstyle{{2(n - 2)\gamma p} \over {2p - 1}}}}}{q^{2p}}}}{{\Pi r_ + ^{{\textstyle{{2p(n - 2)(1 - \gamma ) + 1} \over {2p - 1}}}}}}} \right),
 \end{equation}

\begin{equation}  \label{2.11}
 {\kappa _c} =  - \frac{1}{2}{\left. {\frac{{df(r)}}{{dr}}} \right|_{r = {r_c}}} =  - \frac{{(1 + {\alpha ^2})}}{2}\left( {\frac{{k(n - 2){b^{ - 2\gamma }}}}{{(1 - {\alpha ^2})}}r_c^{2\gamma  - 1} - \frac{{2\Lambda {b^{2\gamma }}}}{{n - 1}}r_c^{1 - 2\gamma } - \frac{{{2^p}p(2p - 1){b^{ - {\textstyle{{2(n - 2)\gamma p} \over {2p - 1}}}}}{q^{2p}}}}{{\Pi r_c^{{\textstyle{{2p(n - 2)(1 - \gamma ) + 1} \over {2p - 1}}}}}}} \right),
 \end{equation}
from which the radiation temperatures of the two horizons can be got by $T_{+,c}={\kappa _{+,c}}/{4\pi}$.

When $f({r_{ + /c}}) = 0$ , we have

 \begin{equation}  \label{2.12}
 \begin{aligned}
 m({r_ + }) =  - Ar_ + ^{{\textstyle{{{\alpha ^2} + n - 2} \over {{\alpha ^2} + 1}}}} + B{q^{2p}}r_ + ^{ - {\textstyle{{{\alpha ^2} - 2p + n} \over {(2p - 1)({\alpha ^2} + 1)}}}} + C\Lambda r_ + ^{ - {\textstyle{{{\alpha ^2} - n} \over {{\alpha ^2} + 1}}}},\\
 m({r_c}) =  - Ar_c^{{\textstyle{{{\alpha ^2} + n - 2} \over {{\alpha ^2} + 1}}}} + B{q^{2p}}r_c^{ - {\textstyle{{{\alpha ^2} - 2p + n} \over {(2p - 1)({\alpha ^2} + 1)}}}} + C\Lambda r_c^{ - {\textstyle{{{\alpha ^2} - n} \over {{\alpha ^2} + 1}}}}.
 \end{aligned}
 \end{equation}

$m({r_ + })=m({r_ c })=m$, and take $x = {r_+ }/{r_c}$  as the ratio of positions of BEH and CEH, which meets $0 < x \le 1$, then

 \begin{equation}  \label{2.13}
 C\Lambda  = A\frac{{r_c^{{\textstyle{{2({\alpha ^2} - 1)} \over {{\alpha ^2} + 1}}}}(1 - {x^{{\textstyle{{{\alpha ^2} + n - 2} \over {{\alpha ^2} + 1}}}}})}}{{(1 - {x^{ - {\textstyle{{{\alpha ^2} - n} \over {{\alpha ^2} + 1}}}}})}} - B{q^{2p}}\frac{{r_c^{{\textstyle{{2p(1 + {\alpha ^2} - n) - 2{\alpha ^2}} \over {(2p - 1)({\alpha ^2} + 1)}}}}(1 - {x^{ - {\textstyle{{{\alpha ^2} - 2p + n} \over {(2p - 1)({\alpha ^2} + 1)}}}}})}}{{(1 - {x^{ - {\textstyle{{{\alpha ^2} - n} \over {{\alpha ^2} + 1}}}}})}},
  \end{equation}

 \begin{equation}  \label{2.14}
 m = Ar_c^{{\textstyle{{{\alpha ^2} + n - 2} \over {{\alpha ^2} + 1}}}}\frac{{{x^{ - {\textstyle{{{\alpha ^2} - n} \over {{\alpha ^2} + 1}}}}} - {x^{{\textstyle{{{\alpha ^2} + n - 2} \over {{\alpha ^2} + 1}}}}}}}{{(1 - {x^{ - {\textstyle{{{\alpha ^2} - n} \over {{\alpha ^2} + 1}}}}})}} + B{q^{2p}}r_c^{ - {\textstyle{{{\alpha ^2} - 2p + n} \over {(2p - 1)({\alpha ^2} + 1)}}}}\frac{{{x^{ - {\textstyle{{{\alpha ^2} - 2p + n} \over {(2p - 1)({\alpha ^2} + 1)}}}}} - {x^{ - {\textstyle{{{\alpha ^2} - n} \over {{\alpha ^2} + 1}}}}}}}{{(1 - {x^{ - {\textstyle{{{\alpha ^2} - n} \over {{\alpha ^2} + 1}}}}})}}.
  \end{equation}

The thermodynamic volumes corresponding to the two horizons  are respectively given by
\begin{equation}  \label{2.15+}
\begin{aligned}
{V_+} = \frac{{({\alpha ^2} + 1){b^{\gamma (n + 1)}}{\omega _{n - 1}}}}{{n - {\alpha ^2}}}r_+^{{\textstyle{{n - {\alpha ^2}} \over {{\alpha ^2} + 1}}}} ,\\
{V_c} = \frac{{({\alpha ^2} + 1){b^{\gamma (n + 1)}}{\omega _{n - 1}}}}{{n - {\alpha ^2}}}r_c^{{\textstyle{{n - {\alpha ^2}} \over {{\alpha ^2} + 1}}}},
\end{aligned}
\end{equation}
where  ${\omega _{n - 1}}$ represents the volume of constant curvature hypersurface described by $d\Omega _{k,n - 1}^2$.

The entropies of BEH and CEH in dS space are expressed respectively as

 \begin{equation}  \label{2.16+}
 \begin{aligned}
 &{S_+}= \frac{{{b^{(n - 1)\gamma }}r_+^{(n - 1)(1 - \gamma )}}}{4}, \\
 &{S_c}= \frac{{{b^{(n - 1)\gamma }}r_c^{(n - 1)(1 - \gamma )}}}{4}.
 \end{aligned}
 \end{equation}

\section{Effective thermodynamics and modified entropy of the EPMD black holes in  dS Space-time}
In general, the radiation temperatures of the BEH and CEH are different. So, If one investigates the black hole in dS space-time including BEH and CEH as a whole thermodynamic system, it is usually thermodynamically unstable or non-equilibrium. We find that the radiation temperatures are equal if the charge of the system satisfies some condition. Under the condition, considering the correlation of the two horizons, we derived the effective thermodynamic quantities and the modified entropies of the EPMD black holes in dS space-time.

When the radiation temperatures of BEH and CEH are equivalent, ${\kappa _ + } = {\kappa _c} $, Eq.(\ref{2.10}) and Eq.(\ref{2.11}) with Eq.(\ref{2.13}) give the condition about the charge for the same radiation temperature of BEH and CEH.

\begin{equation}  \label{2.17}
\frac{{{2^p}p{q^{2p}}(2p - 1){b^{ - {\textstyle{{2(n - 4)p\gamma  + 2\gamma } \over {(2p - 1)}}}}}r_c^{{\textstyle{{2p(3 - n - {\alpha ^2}) - 2} \over {(2p - 1)({\alpha ^2} + 1)}}}}}}{\Pi } = \frac{{k{A_1}(x)}}{{{B_1}(x)}}\frac{{(n - 2)}}{{({\alpha ^2} - 1)}},
 \end{equation}
where
 \begin{equation}  \label{2.18}
 \begin{aligned}
 &{A_1}(x) =  {\frac{{({\alpha ^2} - n)}}{{({\alpha ^2} + n - 2)}}(1 - {x^{{\textstyle{{{\alpha ^2} + n - 2} \over {{\alpha ^2} + 1}}}}})(1 + {x^{\frac{{1 - {\alpha ^2}}}{{{\alpha ^2} + 1}}}}) + (1 + {x^{{\textstyle{{{\alpha ^2} - 1} \over {{\alpha ^2} + 1}}}}})(1 - {x^{ - {\textstyle{{{\alpha ^2} - n} \over {{\alpha ^2} + 1}}}}})},\\
 &{B_1}(x) =  {\frac{{({\alpha ^2} - n)(2p - 1)}}{{({\alpha ^2} + n - 2p)}}(1 + {x^{\frac{{1 - {\alpha ^2}}}{{{\alpha ^2} + 1}}}})(1 - {x^{ - {\textstyle{{{\alpha ^2} - 2p + n} \over {(2p - 1)({\alpha ^2} + 1)}}}}}) - (1 + {x^{ - {\textstyle{{2p(n - 2)(1 - \gamma ) + 1} \over {2p - 1}}}}})(1 - {x^{ - {\textstyle{{{\alpha ^2} - n} \over {{\alpha ^2} + 1}}}}})}.
 \end{aligned}
 \end{equation}
Substituting Eq.(\ref{2.13}) and Eq.(\ref{2.17}) into Eq.(\ref{2.11}), the temperature $T$ for the same radiation temperature of BEH and CEH can be obtained as

\begin{equation}  \label{2.19}
\begin{aligned}
&T = {T_ + } = {T_c}=-\frac{{(1 + {\alpha ^2})}}{{4\pi }}kr_c^{{\textstyle{{({\alpha ^2} - 1)} \over {{\alpha ^2} + 1}}}}{b^{ - 2\gamma }}\frac{{(n - 2)}}{{(1 - {\alpha ^2})}}\\
&\left\{ {\left[ {1 + \frac{{({\alpha ^2} - n)}}{{({\alpha ^2} + n - 2)}}\frac{{(1 - {x^{{\textstyle{{{\alpha ^2} + n - 2} \over {{\alpha ^2} + 1}}}}})}}{{(1 - {x^{ - {\textstyle{{{\alpha ^2} - n} \over {{\alpha ^2} + 1}}}}})}}} \right]} \right. - \frac{{{A_1}(x)}}{{{B_1}(x)}}\left. {\left[ {1 - \frac{{({\alpha ^2} - n)(2p - 1)}}{{({\alpha ^2} + n - 2p)}}\frac{{(1 - {x^{ - {\textstyle{{{\alpha ^2} - 2p + n} \over {(2p - 1)({\alpha ^2} + 1)}}}}})}}{{(1 - {x^{ - {\textstyle{{{\alpha ^2} - n} \over {{\alpha ^2} + 1}}}}})}}} \right]} \right\}.
\end{aligned}
\end{equation}

Substituting Eq.(\ref{2.17}) into Eq.(\ref{2.14}) and Eq.(\ref{2.8}), it gives the energy(mass) of the EPMD black holes in dS space-time as

 \begin{equation}  \label{3.1}
 M = \frac{{{b^{(n - 1)\gamma }}(n - 1)}}{{16\pi ({\alpha ^2} + 1)}}r_c^{{\textstyle{{{\alpha ^2} + n - 2} \over {{\alpha ^2} + 1}}}}\left[ {A(x) + {q^{2p}}r_c^{ - {\textstyle{{2p(n + {\alpha ^2} - 3) + 2} \over {(2p - 1)({\alpha ^2} + 1)}}}}B(x)} \right],
 \end{equation}
where
 \begin{equation}  \label{3.2}
 A(x) = A\frac{{{x^{ - {\textstyle{{{\alpha ^2} - n} \over {{\alpha ^2} + 1}}}}} - {x^{{\textstyle{{{\alpha ^2} + n - 2} \over {{\alpha ^2} + 1}}}}}}}{{(1 - {x^{ - {\textstyle{{{\alpha ^2} - n} \over {{\alpha ^2} + 1}}}}})}}, B(x) = B\frac{{{x^{ - {\textstyle{{{\alpha ^2} - 2p + n} \over {(2p - 1)({\alpha ^2} + 1)}}}}} - {x^{ - {\textstyle{{{\alpha ^2} - n} \over {{\alpha ^2} + 1}}}}}}}{{(1 - {x^{ - {\textstyle{{{\alpha ^2} - n} \over {{\alpha ^2} + 1}}}}})}} .
 \end{equation}

Taking the EPMD dS space-time as a thermodynamic system, in Refs. \cite{Saoussen19,Brian13a,ZhangLC16a,Urano09,Dayyani17,Mo16}, the thermodynamic volume of the EPMD dS space-time is given by

\begin{equation}\label{3.3}
V = {V_c} - {V_ + }.
\end{equation}

Considering the correlation of BEH and CEH,  we assume that the entropy of the EPMD dS space-time is expressed as
 \begin{equation}  \label{3.5}
 S =S_{c}+S_{+}+S_{AB}= \frac{{{b^{(n - 1)\gamma }}r_c^{(n - 1)(1 - \gamma )}}}{4}[1 + {x^{(n - 1)(1 - \gamma )}} + {f_{AB}}(x)] = \frac{{{b^{(n - 1)\gamma }}r_c^{(n - 1)(1 - \gamma )}}}{4}{F_n}(x),
 \end{equation}
where ${f_{AB}}(x)$ is an arbitrary function of $x$ .

Using the effective thermodynamic quantities, the state parameters of the thermodynamic system satisfy the formula of the first law of thermodynamics, i.e.,
\begin{equation}  \label{3.6}
dM = {T_{eff}}dS - {P_{eff}}dV + {\Phi _{eff}}dQ ,
\end{equation}
where the effective temperature  ${T_{eff}}$ , the effective pressure  ${P_{eff}}$ and the effective potential  ${\Phi _{eff}}$ of the EPMD dS black hole system are respectively defined as

 \begin{equation}  \label{3.7}
 {T_{eff}} = {\left( {\frac{{\partial M}}{{\partial S}}} \right)_{Q,V}} = \frac{{k(n - 2)(n - 1){b^{ - 2\gamma }}}}{{4\pi ({\alpha ^2} - 1)({\alpha ^2} + n - 2)x(1 - {x^{ - {\textstyle{{{\alpha ^2} - n} \over {{\alpha ^2} + 1}}}}})}}r_c^{{\textstyle{{{\alpha ^2} - 1} \over {{\alpha ^2} + 1}}}}\frac{{{T_3}(x{\rm{)}}}}{{{T_2}(x)}},
 \end{equation}
with
\begin{equation}  \label{3.8}
\begin{aligned}
 {T_3}(x) =&  - \left[ {({\alpha ^2} + n - 2)({x^{{\textstyle{{{\alpha ^2} + n - 2} \over {{\alpha ^2} + 1}}}}} - {x^{{\textstyle{{2n - 2{\alpha ^2}} \over {{\alpha ^2} + 1}}}}}) + ({\alpha ^2} - n){x^{ - {\textstyle{{{\alpha ^2} - n} \over {{\alpha ^2} + 1}}}}}(1 - {x^{{\textstyle{{{\alpha ^2} + n - 2} \over {{\alpha ^2} + 1}}}}})} \right]\\
 &+{q^{2p}}r_c ^{ - {\textstyle{{2p{\alpha ^2} - 6p + 2np + 2} \over {(2p - 1)({\alpha ^2} + 1)}}}}\frac{{{2^p}p{{(2p - 1)}^2}({\alpha ^2} + n - 2)({\alpha ^2} - 1){b^{ - {\textstyle{{2(n - 2)p\gamma } \over {(2p - 1)}}} + 2\gamma }}}}{{\Pi ({\alpha ^2} + n - 2p)(n - 2)k}}\\
 &\left[ { - \frac{{({\alpha ^2} - 2p + n)({x^{ - {\textstyle{{{\alpha ^2} - 2p + n} \over {(2p - 1)({\alpha ^2} + 1)}}}}} - {x^{{\textstyle{{2n - 2{\alpha ^2}} \over {{\alpha ^2} + 1}}}}})}}{{(2p - 1)}} + ({\alpha ^2} - n){x^{ - {\textstyle{{{\alpha ^2} - n} \over {{\alpha ^2} + 1}}}}}(1 - {x^{ - {\textstyle{{{\alpha ^2} - 2p + n} \over {(2p - 1)({\alpha ^2} + 1)}}}}})} \right],
 \end{aligned}
 \end{equation}

\begin{equation}  \label{3.9}
{T_2}(x)
= {F_n}'(x)(1 - {x^{{\textstyle{{n - {\alpha ^2}} \over {{\alpha ^2} + 1}}}}}) + {x^{{\textstyle{{n - 1 - 2{\alpha ^2}} \over {{\alpha ^2} + 1}}}}}(n - 1)(1 - \gamma ){F_n}(x).
\end{equation}

When $T = {T_ + } = {T_c}$, which means the dS black hole thermodynamic system including BEH and CEH are thermodynamically equilibrium, the effective temperature of the space-time should be equal to the radiation temperatures of BEH and CEH so as to it is with thermodynamic significance. So, when the charge of the space-time  ${q^{2p}}$ is expressed by Eq.(\ref{2.17}),  ${T_{eff}} = {\tilde T_{eff}}{\rm{ = }}{T_ + } = {T_c}$. Substituting Eq.(\ref{2.17}) into Eq.(\ref{3.7}), it gives

\begin{equation}  \label{3.10}
\begin{aligned}
{\tilde T_{eff}} = & - \frac{{{{(1 + {\alpha ^2})}^3}}}{{(n - 1)}}k{b^{ - 2\gamma }}\frac{{(n - 2)}}{{(1 - {\alpha ^2})}}\left\{ {\left[ {1 + \frac{{({\alpha ^2} - n)}}{{({\alpha ^2} + n - 2)}}\frac{{(1 - {x^{{\textstyle{{{\alpha ^2} + n - 2} \over {{\alpha ^2} + 1}}}}})}}{{(1 - {x^{ - {\textstyle{{{\alpha ^2} - n} \over {{\alpha ^2} + 1}}}}})}}} \right]} \right.\\
  &- \frac{{{A_1}(x)}}{{{B_1}(x)}}\left. {\left[ {1 - \frac{{({\alpha ^2} - n)(2p - 1)}}{{({\alpha ^2} + n - 2p)}}\frac{{(1 - {x^{ - {\textstyle{{{\alpha ^2} - 2p + n} \over {(2p - 1)({\alpha ^2} + 1)}}}}})}}{{(1 - {x^{ - {\textstyle{{{\alpha ^2} - n} \over {{\alpha ^2} + 1}}}}})}}} \right]} \right\} = \frac{{{{\tilde T}_1}(x)}}{{{T_2}(x)}}.
 \end{aligned}
 \end{equation}
where
\begin{equation}  \label{3.11}
{\tilde T_1}(x) = \frac{{k(n - 2){{({\alpha ^2} + 1)}^2}{b^{ - 2\gamma }}}}{{({\alpha ^2} - 1)x(1 - {x^{ - {\textstyle{{{\alpha ^2} - n} \over {{\alpha ^2} + 1}}}}})}}{\tilde T_3}(x),
 \end{equation}

\begin{equation}  \label{3.12}
 \begin{aligned}
 {\tilde T_3}(x) =&  - ({\alpha ^2} + n - 2)({x^{{\textstyle{{{\alpha ^2} + n - 2} \over {{\alpha ^2} + 1}}}}} - {x^{{\textstyle{{2n - 2{\alpha ^2}} \over {{\alpha ^2} + 1}}}}}) - ({\alpha ^2} - n){x^{ - {\textstyle{{{\alpha ^2} - n} \over {{\alpha ^2} + 1}}}}}(1 - {x^{{\textstyle{{{\alpha ^2} + n - 2} \over {{\alpha ^2} + 1}}}}}) + \frac{{{A_1}(x)}}{{{B_1}(x)}}\frac{{({\alpha ^2} + n - 2)(2p - 1)}}{{({\alpha ^2} + n - 2p)}}\\
  & \left\{ { - \frac{{({\alpha ^2} - 2p + n)[{x^{ - {\textstyle{{{\alpha ^2} - 2p + n} \over {(2p - 1)({\alpha ^2} + 1)}}}}} - {x^{{\textstyle{{2n - 2{\alpha ^2}} \over {{\alpha ^2} + 1}}}}}]}}{{(2p - 1)}} + ({\alpha ^2} - n){x^{ - {\textstyle{{{\alpha ^2} - n} \over {{\alpha ^2} + 1}}}}}[1 - {x^{ - {\textstyle{{{\alpha ^2} - 2p + n} \over {(2p - 1)({\alpha ^2} + 1)}}}}}]} \right\}.
\end{aligned}
\end{equation}
Comparing Eq.(\ref{3.10}) with Eq.(\ref{2.19}), one can get

 \begin{equation}  \label{3.13}
 {T_2}(x) = \frac{{(n - 1)({x^{{\textstyle{{n - 1} \over {{\alpha ^2} + 1}}}}} + {x^{{\textstyle{{2n - 2{\alpha ^2}} \over {{\alpha ^2} + 1}}}}})}}{{(1 + {\alpha ^2})x(1 - {x^{ - {\textstyle{{{\alpha ^2} - n} \over {{\alpha ^2} + 1}}}}})}} = \frac{{(n - 1)(1 - \gamma ){x^{n - 2 - n\gamma  + \gamma }}(1 + {x^{n + 1 - n\gamma  - 3\gamma }})}}{{(1 - {x^{n - \gamma (n + 1)}})}}.
 \end{equation}
From Eq.(\ref{3.13}) and Eq.(\ref{3.9}), a differential equation of ${F_n}(x)$  can be obtained. Taking the initial condition as ${F_n}(0) = 1$, ie., ${f_{AB}}(0)=0$, which indicates that the interaction between the two horizons is zero when $x=0$, the solutions of the differential equation are
\begin{equation}  \label{3.14}
\begin{aligned}
{F_n}(x)
&= \frac{{3(n - \gamma n) - \gamma  - 1}}{{2(n - \gamma n) - 1}}{[1 - {x^{(n - \gamma n - \gamma )}}]^{(n - 1)(1 - \gamma )/(n - \gamma n - \gamma )}}\\
&\quad -\frac{{(n - \gamma n - \gamma )[1 + {x^{2(n - \gamma n) - 1}}] + [2(n - \gamma n) - 1](1 - 2{x^{n - n\gamma  - \gamma }} - {x^{2n - 2n\gamma  - 1}})}}{{[2(n - \gamma n) - 1][1 - {x^{(n - \gamma n - \gamma )}}]}} + 1 + {x^{(n - 1)(1 - \gamma )}}\\
& = {f_{AB}}(x) + 1 + {x^{(n - 1)(1 - \gamma )}}.
\end{aligned}
\end{equation}
Substituting  Eq.(\ref{3.14}) into Eq.(\ref{3.7}), the effective temperature of the EPMD dS space-time can be expressed as

 \begin{equation}  \label{3.15}
 {T_{eff}} = \frac{{k(n - 2){b^{ - 2\gamma }}{T_3}(x{\rm{)}}}}{{4\pi r_c^{1 - 2\gamma }({\alpha ^2} - 1)({\alpha ^2} + n - 2)(1 - \gamma ){x^{n - 1 - n\gamma  + \gamma }}(1 + {x^{n + 1 - n\gamma  - 3\gamma }})}}.
 \end{equation}

\section{Entropic force between the two horizons in the EPMD dS space-time}

The definition of the entropic force in the thermodynamic system is \cite{Plastino18,Zamora18,Panos19,Calderon19,Plastino19a,Plastino19b,Plastino19c,Plastino19d,Nobuyoshi19,Nobuyoshi17,Nobuyoshi16,Cai11a,Cai10a,Treumann19}
\begin{equation}  \label{4.1}
F =  - T\frac{{\partial S}}{{\partial r}} ,
\end{equation}
where $T$ is the system temperature and $r$ is the system radius. From Eq.(\ref{3.5}), the entropy created by the interaction between BEH and CEH is
 \begin{equation}  \label{4.2}
 {S_{AB}} = \frac{{{b^{(n - 1)\gamma }}r_c^{(n - 1)(1 - \gamma )}}}{4}{f_{AB}}(x).
 \end{equation}
According to the expression Eq.(\ref{4.1}), the corresponding entropic force between the two horizons can be given as

 \begin{equation}  \label{4.3}
 F = -{T_{eff}}{\left( {\frac{{\partial {S_{AB}}}}{{\partial r}}} \right)_{{T_{eff}}}},
 \end{equation}
where  ${T_{eff}}$ is the effective temperature of the system and  $r = {r_c} - {r_ + } = {r_c}(1 - x)$. Then

\begin{equation}  \label{4.4}
\begin{aligned}
F(x)=& -\frac{{k(n - 2){b^{(n - 3)\gamma }}r_c^{(n - 3)(1 - \gamma )}{T_3}(x{\rm{)}}}}{{16\pi ({\alpha ^2} - 1)({\alpha ^2} + n - 2)(1 - \gamma )}}\\
& \frac{{(n - 1)(1 - \gamma ){f_{AB}}(x)\frac{d}{{dx}}\left[ {\frac{{{T_3}(x{\rm{)}}}}{{{x^{n - 1 - n\gamma  + \gamma }}(1 + {x^{n + 1 - n\gamma  - 3\gamma }})}}} \right] + \frac{{{T_3}(x{\rm{)}}}{f_{AB}}'(x)}{{{x^{n - 1 - n\gamma  + \gamma }}(1 + {x^{n + 1 - n\gamma  - 3\gamma }})}}}}{{{x^{n - 1 - n\gamma  + \gamma }}(1 + {x^{n + 1 - n\gamma  - 3\gamma }})(1 - x)\frac{d}{{dx}}\left[ {\frac{{{T_3}(x{\rm{)}}}}{{{x^{n - 1 - n\gamma  + \gamma }}(1 + {x^{n + 1 - n\gamma  - 3\gamma }})}}} \right] - {T_3}(x{\rm{)}}}}.
\end{aligned}
 \end{equation}

In order to describe the behaviors of the entropic force created by the interaction between BEH and CEH and the effect of the parameters of the MPMD dS space-time on the entropic force, the solutions for the entropic force $F(x)$ with different parameters, $n$, $\alpha$, $p$ and $\kappa$ are depicted in the following figures, where we have taken ${q^{2p}}r_ c ^{ - {\textstyle{{2p(n + {\alpha ^2} - 3) + 2} \over {(2p - 1)({\alpha ^2} + 1)}}}}{b^{ - {\textstyle{{2(n - 2)p\gamma } \over {(2p - 1)}}} + 2\gamma }}=\kappa$, $\frac{{k(n - 2){b^{(n - 3)\gamma }}r_c^{(n - 3)(1 - \gamma )}}}{{16\pi ({\alpha ^2} - 1)({\alpha ^2} + n - 2)(1 - \gamma )}} = 1$  and $k = 1$.

\begin{figure}[htp]
\begin{minipage}[t]{0.45\textwidth}
  \centering
  \includegraphics[width=3in]{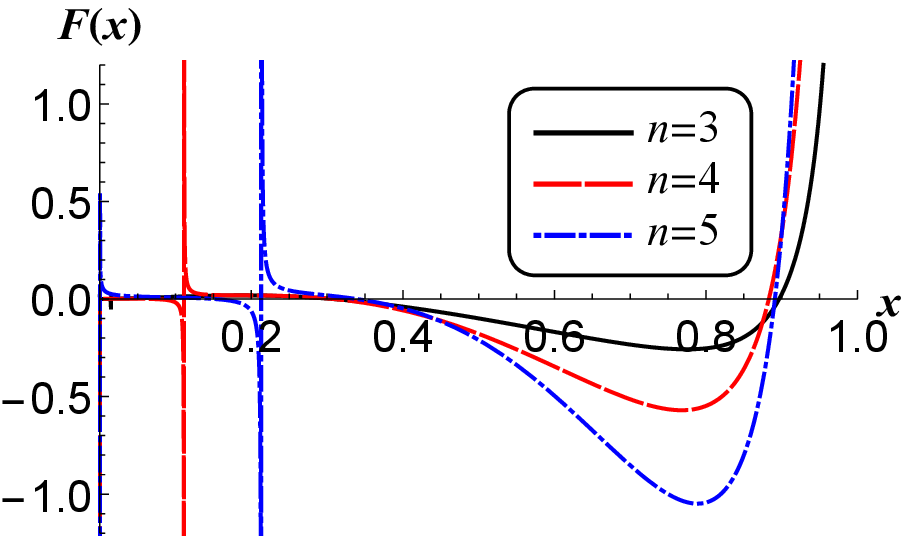}
  \caption{$F(x)-x $ curve with different values of the parameter $n$ for $\alpha=0.3$, $p=1.3$ and $\kappa=0.001$.}\label{fig1}
\end{minipage}
\begin{minipage}[t]{0.45\textwidth}
  \centering
  \includegraphics[width=3in]{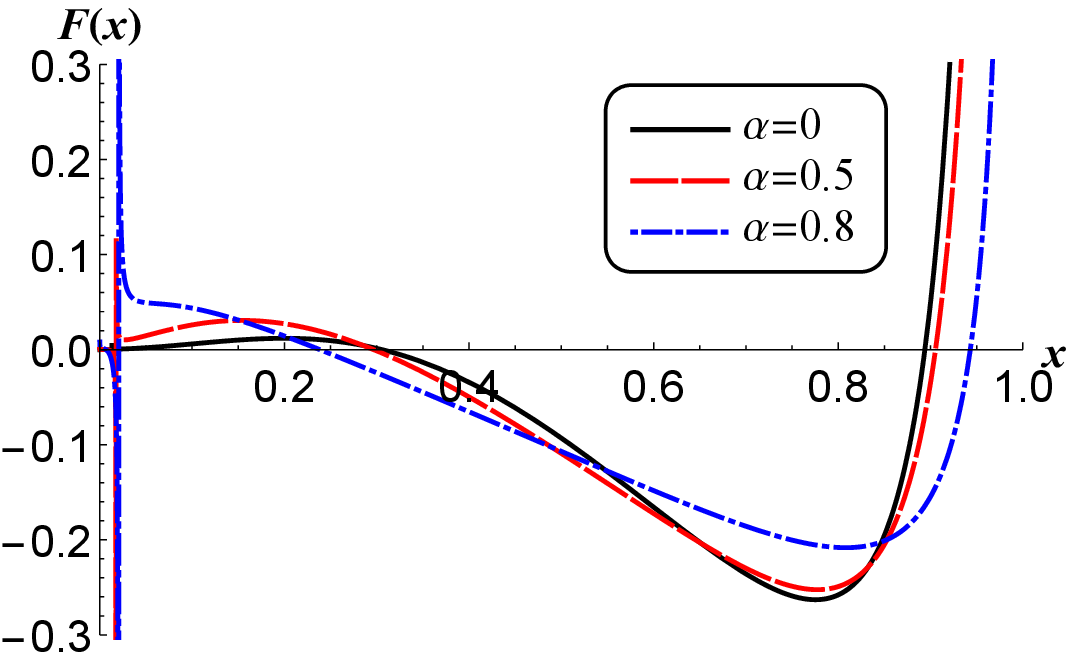}
  \caption{$F(x)-x $ curve with different values of the parameter $\alpha$ for $n=3$, $p=1.3$ and $\kappa=0.001$.}\label{fig2}
\end{minipage}
\end{figure}

\begin{figure}[htp]
\begin{minipage}[t]{0.45\textwidth}
  \centering
  \includegraphics[width=3in]{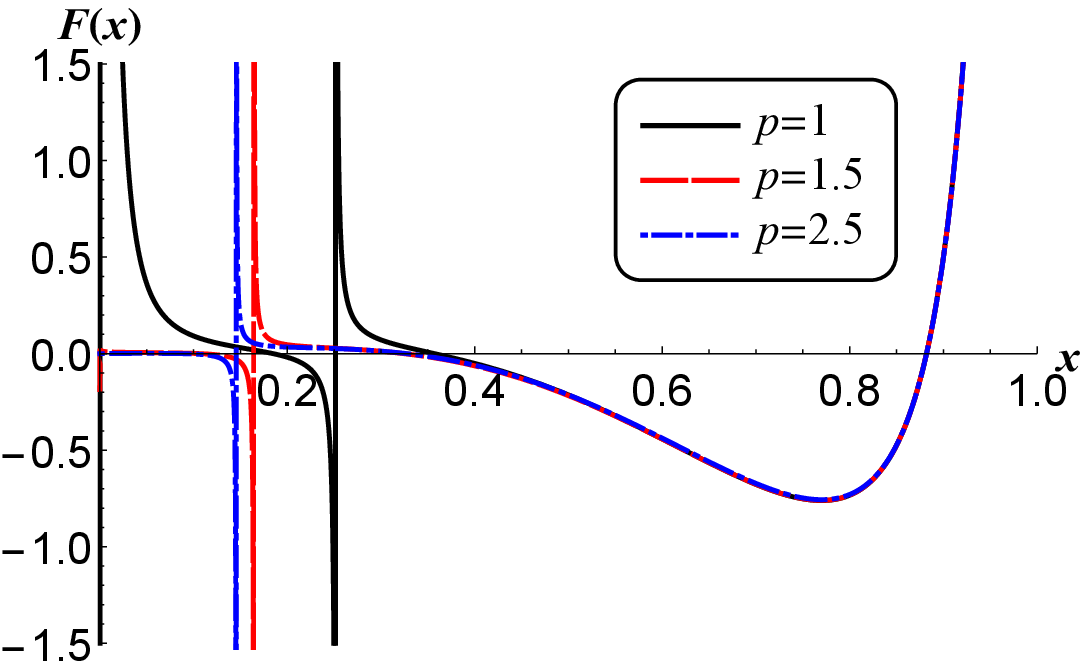}
  \caption{$F(x)-x $ curve with different values of the parameter $p$ for $n=5$, $\alpha=0.5$ and $\kappa=0.001$.}\label{fig3}
\end{minipage}
\begin{minipage}[t]{0.45\textwidth}
  \centering
  \includegraphics[width=3in]{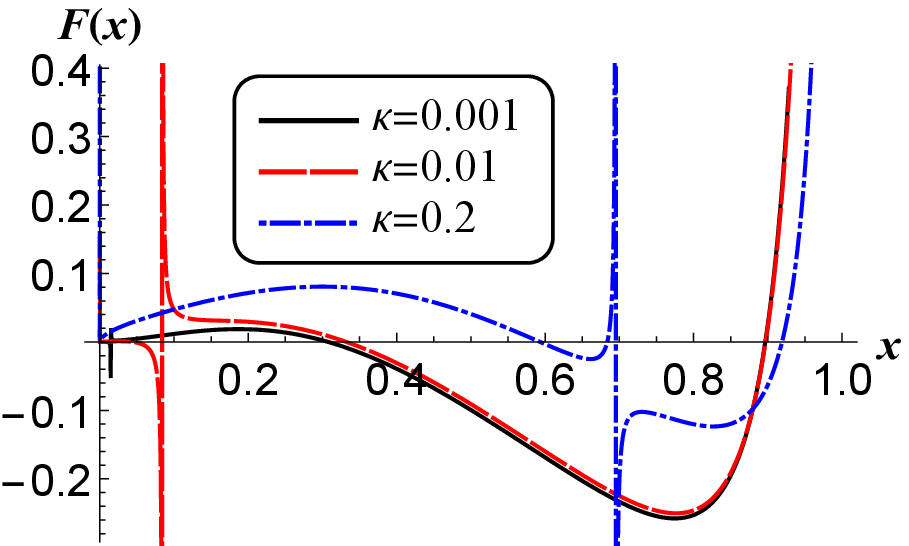}
  \caption{$ F(x)-x $ curve with different values of the parameter $\kappa$ for $n=3$, $\alpha=0.3$ and $p=1.3$.}\label{fig4}
\end{minipage}
\end{figure}

If the effect of entropic force is same to that of normal forces, from the $F(x)-x$  curves, it is clear that the entropic force tends to infinity with $x \to 1$ , which means that when the BEH is close to the CEH in the dS space-time, the two horizons will separate from each other due to the entropic force with a corresponding large acceleration provided the other forces are absent. This agrees with the present viewpoint on early cosmic inflation. When the value of $x$ reduces from 1, the entropic force between the two horizons is decreasing until it reaches a minimum, and at $x = {x_1}$ in the interval, the entropic force is zero,  which can be interpreted as that the interaction between the two horizons is absent at $x = {x_1}$ , however where the two horizons may keep separation state. When the value of $x$ reduces gradually, the value of the entropic force keeps in a negative territory temporarily, which means that the separation of the two horizons is decelerated and can be interpreted as the cosmic decelerated expansion. If the expansion speed is decelerated to zero before $x$ reach to minimum at $x={x_2}$, the two horizons will be in a relative oscillatory motion with the equilibrium position $x={x_1}$ under the circumstance of no other forces exist. In the four $F(x)-x$ figures, when $x$ reduces from $x={x_2}$, the value of the entropic force tends to zero in the negative territory before it goes to positive value (meaning repulsive force) at $x={x_3}$ except the blue dotdashed curve in FIG.4. In the region of smaller $x$, the behaviors of the entropic force are complicated, most of curves in the four $F(x)-x$ figures go through their singularities and go from positive to negative with decreasing $x$. but the singularity disappears on the  black solid curves in FIG.2, which corresponds to the situations of smaller $n$ with $n=3$, smaller $\alpha$ with $\alpha=0$, smaller $p$ with $p=1.3$, and smaller $\kappa$ with $\kappa=0.001$. Besides that, when $\kappa$ is bigger the behavior of the entropic force is different, which can be seen from the blue dotdashed curve in FIG.4. These situations of $F(x)-x$ curves indicate that the behavior of entropic force is affected by the parameters $n$, $\alpha$, $p$ and $\kappa$, that is, it is influenced by the dimension of the space-time, the nonlinearity of the electromagnetic field, the strength of coupling of the dilaton scalar and electromagnetic field, the position of cosmological horizon, and the electric charge of the black hole. In all these cases of the four figures, the behaviors of the entropic force near $x={x_1}$, or in the region of $1>x>x_3$ are similar to that of Lennard-Jones force between two particles \cite{Zhang19,Miao18,Miao19}. They are similar but obtained by completely different ways. This indicates that the entropic force between the two horizons has a certain internal relationship with Lennard-Jones force between two particles. What is the fate of the accelerated expanding universe, whether the entropic force between BEH and CEH is one of the participant forces which drive the evolution of the universe, and whether the entropic force has the same effect with Lennard-Jones force, which need more investigations and more evidences.

\section{Conclusion and discussion}
The entropy of the charged dilaton black holes with Einstein power-Maxwell field in dS space-time is derived and discussed in the paper considering the correlation between BEH and CEH, especially the correction term caused by the interaction between BEH and CEH. The entropic force $F(x)$ between BEH and CEH is deduced according to the definition of the entropic force in thermodynamic system. We discuss the entropic force $F(x)$ changes with $x$, the position ratio of BEH and CEH, in the EPMD dS space-time when the the parameters $n$, $\alpha$, $p$, and $\kappa$ take some certain values. It is found that the behaviors of the entropic force $F(x)$  at a larger interval of $x$ are similar to that of Lennard-Jones force between two particles, and in a smaller interval of $x$ the behaviors of the entropic force $F(x)$ are complicated, which are related to the parameters of the space-time.

Comparing the $F(x) -x $ curves in a large interval of $x$ with the curve of Lennard-Jones force versus the distance of two particles given in reference\cite{Zhang19,Miao18,Miao19}, we find that the two curves are very similar although they are obtained in different ways. The entropic force between the two horizons is completely derived from general relativity. But the Lennard-Jones force between two particles is concluded from simulation based on experiments. This indicates that there may be a relationship between the entropic force and the Lennard-Jones force.

According to modern cosmology, the fate of our universe is dominated by matter and energy. If there are enough matter and energy, the gravitational effect will stop the cosmic expansion at a certain time, and then our universe will turn to contraction. Otherwise, if the density of cosmic matter and energy is too low, the universe will expand forever.  Different kinds of cosmic matter and energy play different roles in the cosmic expansion. In this work, we find that the entropic force probably plays a certain role in the cosmic expansion and contraction, other than cosmic matter and energy. In the EPMD dS space-time, the influence of different parameters on the entropic force between BEH and CEH is shown in the $F(x) - x$ curves. This indicates that different parameters in the EPMD dS space-time play different roles in the cosmic expansion.

Since the space-time and the thermodynamic effect are relative to general relativity and quantum mechanics respectively, and all physical quantities satisfy the first law of thermodynamics. If the effect of entropyic force is proved similar to that of normal forces, it indicates that there is a relationship among general relativity, quantum mechanics and thermodynamics. It will provide a new way to study the interaction between particles in black holes, the microstate of particles in black holes, the Lennard-Jones potential between particles and the microstate of particles in ordinary thermodynamic systems.

\section*{Acknowledgments}

We thank Prof. Z. H. Zhu for useful discussions.

This work was supported by the National Natural Science Foundation of China (Grant Nos.11705107, 11847123, 11475108, 11705106, 11605107), the Natural Science Foundation of Shanxi Province, China (Grant No. 201601D102004), the Scientific and Technological Innovation Programs of Higher Education Institutions of Shanxi Province, China (Grant No. 2019L0743, No. 2020L0471, No. 2020L0472).

\end{document}